# DEEPSPACE: DYNAMIC SPATIAL AND SOURCE CUE BASED SOURCE SEPARATION FOR DIALOG ENHANCEMENT

*Aaron Master\*, Lie Lu\*, Jonas Samuelsson\*\*, Heidi-Maria Lehtonen\*\*, Scott Norcross\*,*
*Nathan Swedlow\*, and Audrey Howard\**

\*Dolby Laboratories, Inc, 1275 Market St, San Francisco, CA 94103, United States
\*\*Dolby Sweden AB, Gävlegatan 12A, 113 30 Stockholm, Sweden

## ABSTRACT

Dialog Enhancement (DE) is a feature which allows a user to increase the level of dialog in TV or movie content relative to non-dialog sounds. When only the original mix is available, DE is "unguided," and requires source separation. In this paper, we describe the DeepSpace system, which performs source separation using both dynamic spatial cues and source cues to support unguided DE. Its technologies include spatio-level filtering (SLF) and deep-learning based dialog classification and denoising. Using subjective listening tests, we show that DeepSpace demonstrates significantly improved overall performance relative to state-of-the-art systems available for testing. We explore the feasibility of using existing automated metrics to evaluate unguided DE systems.

## 1. INTRODUCTION

Dialog enhancement (DE) (see, e.g. [1, 2]) is a feature typically used for TV and movie content which allows users to increase the level of dialog relative to other sounds, typically referred to as backgrounds or music and effects ("M&E"). DE can address issues with intelligibility [3] or simply increase user satisfaction [4]. Here, we define a *dialog-boosted signal*, or *DE output signal* as:

$$y = gd + x \qquad (1)$$

where $x$ is the original input mix signal (consisting of dialog $d$ plus backgrounds $b$), and $g$ is the dialog boost factor derived from listener or other input. If ground truth clean dialog $d$ is available, then $y$ is the *ideal dialog-boosted signal*. In that case, $y = (g+1)d + b$ and the boost gain $g$ may be derived from the desired decibel increase in dialog $g_{dB}$ as $g = 10^{\wedge}(g_{dB}/20) - 1$. (Scaling may also be applied to $y$ such that the boosted signal preserves the loudness or level of $d$ or $y$.) Presently, we consider the *unguided* DE case where $d$ is unavailable, must be estimated via source separation (see, e.g., [5, 6]), and shall be termed $\hat{d}$, with $y$ becoming the *estimated dialog-boosted signal*. This paper presents a source separation method, termed as DeepSpace (DS), to estimate $\hat{d}$ using both dynamic spatial and source cues.

DE as defined here differs in key ways from *speech enhancement* (SE) (or, equivalently, *speech denoising*). DE deliberately includes backgrounds (see Eqn. 1) while SE aims to completely eliminate backgrounds. (In practical implementations, SE may include some of the original mix, but evaluations of SE typically do not.) As DE includes the original mix, the source separation used to support it aims to prevent distortion when adding estimated dialog back to the original mix. SE systems are typically designed for voice communications and use monaural audio and sample rates of 16 kHz (though some use 48 kHz), while the most common formats for TV and movie content processed by DE are stereo and 5.1, each with 48 kHz sampling. DE systems therefore aim to produce spatially pleasing, wideband audio.

Recent state-of-the-art SE systems include FullSubnet [7], DPCRN [8], and NSNet2 [9], official baseline for the Deep Noise Suppression challenge [10]. However, it is suboptimal to directly apply them to DE since they largely ignore spatial cues and often cannot produce wideband audio. Recent DE systems include [2] which uses both source cues and some spatial cues, [11] which uses deep learning methods, and [1], a predecessor system to DeepSpace which uses dynamic spatial cues and some source cues. The DeepSpace system proposed here offers robust performance by using both dynamic spatial cues and deep learning-based SE systems which synergistically exploit source cues.

There are additional practical considerations for DE. Stereo is more challenging for DE than 5.1 since it has fewer channels from which to exploit spatial information; for stereo, the number of sources (including dialog sources) is frequently larger than the number of channels (two), effectively presenting an under-determined source separation problem. Therefore, in this paper, we focus on DeepSpace source separation for stereo, 48 kHz sampled audio. In practical application to entertainment content, a DE system may exist at various locations in the content delivery chain, leading to various requirements regarding input channel format, input encoding status, or available computation, memory, or latency. We will describe a version of DeepSpace that aims for high quality with limited latency on unencoded or encoded-then-decoded inputs, with computation and memory requirements that are currently feasible for processing at encoding or in cloud-based workflows.

The remainder of this paper is organized as follows. In section 2, we describe motivation for and operation of the DeepSpace system, including SLF, deep learning-based SE, deep-learning based classification and gating, and the architecture used to combine these technologies. In section 3 we present results from subjective listening tests and demonstrate significantly improved overall performance of DeepSpace relative to other systems. In section 4, we explore the feasibility of using automated metrics to evaluate dialog-boosted signal quality. We also present data for a larger set of content items when using DeepSpace as an SE system. In section 5, we conclude with a discussion and areas for future work.

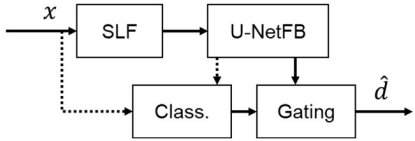

Figure 1: DeepSpace system architecture.

## 2. DEEPSPACE SYSTEM

The DeepSpace system combines SLF with deep learning-based SE and deep learning-based speech classification ("C"), in which SLF primarily exploits spatial cues while SE and C exploit source cues.

Presently, we summarize the predecessor system to DeepSpace, termed "SLF+C", described in detail in [1]. SLF extracts signals whose spatial and level features make them likely dialog candidates, based on characteristics of idiomatically mixed dialog and backgrounds in TV and movie content (See Sec. 3 in [1]). This includes dialog which is panned between the channels (including center-panned and non-center panned) or mixed with reverberation, interchannel phase offset, and/or interchannel delay, as well as dialog whose mixing changes versus time. SLF is thus more spatially robust and dynamic than a system which extracts center-panned dialog as in [12] or which biases in favor of center dialog using a mapping based on interchannel level difference [2]. SLF estimates the mixing parameters of signals which are spatially concentrated vs time (over a 10-frame buffer, hopped every 5 frames, each frame 4096 samples long and a 1024 sample frame hop size for a 48 kHz sample rate) and frequency (over approximately octave-width subbands) and then uses a Bayesian method to estimate a softmask for signals with the identified mixing parameters. A deep learning-based dialog classifier trained on source cues (also described in [1]) receives input of either the original mix or the SLF output, and outputs a binary decision versus time on whether the signal contains dialog, which is then converted to a gating function with ramps, to be applied to the SLF system output. This approach was effective on the majority of tested TV and movie content but exhibited strain to the extent that backgrounds are spatially similar to, and co-occur in time with, dialog. For instance, if dialog is mixed center-left, and engine noise is mixed identically center-left at the same time, then some engine noise could erroneously be included in the dialog estimate.

To address such cases, DeepSpace adds its own SE system which makes greater use of speech source cues than SLF+C. This SE network uses a U-Net type architecture, conceptually similar to [13, 14, 15, 16] but where the inputs are frequency band energies, rather than STFT bin values, and the outputs are real-valued frequency band softmask values; hence we term the system U-NetFB. Given stereo input, U-NetFB downmixes the signal before processing, and applies the output softmask to each input channel.

For the particular synergy between SLF and U-NetFB systems, focused listening exercises found it optimal to combine technologies via the configuration shown in figure 1, and with SLF tuned to be less aggressive in suppressing backgrounds than in [1] where it works without an SE system. U-NetFB receives the SLF output signal as input, and produces a denoised output, which is then gated based on the classifier. Similar to [1], the classifier may take as input either the input mix or the processed signal (each option shown in the figure via dotted lines), with the former allowing less latency (the maximum of SLF+U-NetFB and classifier latencies) than the latter (the sum of all these latencies), with a possible risk of less accuracy. Here, we favor lower latency and use the mix as the classifier input.

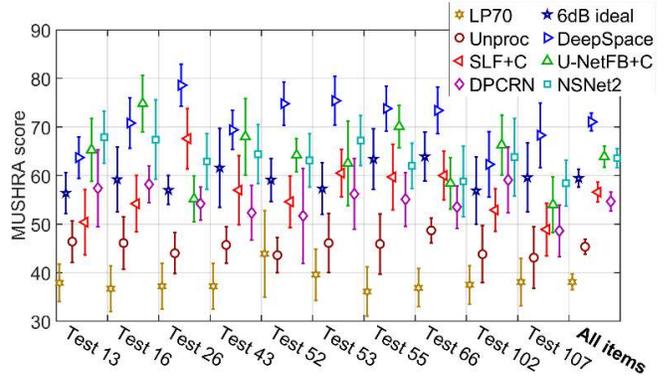

Figure 2: MUSHRA test results

## 3. SUBJECTIVE EVALUATION

We evaluate the DeepSpace system using three subjective listening tests – a MUSHRA test and two absolute category rating (ACR) tests – which present dialog-boosted signals and other related signals to expert human listeners over headphones. None of the test items were used for development of DeepSpace or its components.

### 3.1 MUSHRA Test

MUSHRA tests [17] allow evaluation of systems under test (SUTs) against an ideal reference, on a scale from 0 to 100 with 100 representing a test signal indistinguishable from the reference. In this case we must have clean dialog tracks to generate the reference ideal *dialog-boosted signal*. For this purpose, we choose a large subset of the real-world signals described in the MUSHRA test in [1] which represent a variety of genres, speaker gender presentations, dialog-to-nondialog ratios (DNRs) and panned dialog mixing. For boost level, we chose $g_{dB} = 12$ for all SUTs, as pilot listening found this level to be toward the higher end of the range of preferences across a large number of items with similar DNRs (see also, e.g. [4]); higher boost levels tend to expose more artifacts and lower ones tend to expose fewer.

In this evaluation, we consider the performance of DeepSpace, its components based primarily on spatial cues (SLF+C) and source cues (U-NetFB+C, using the same classifier), and additional systems. Because other DE systems (e.g. [2, 11]) were not available to process our private data, we chose available state of the art SE systems, although they are designed for real-time communications. Based on their relatively strong results in pilot listening tests, we chose DPCRN [8] and NSNet2 [9] rather than FullSubnet [7]. For each, we processed the two stereo channels separately; and for DPCRN, we performed appropriate downsampling and upsampling. For anchors, we used the traditional MUSHRA lowpass anchors, and two others: an unprocessed input mix signal ($g_{dB} = 0$) and an ideal dialog boosted signal with only half the boost of the reference ($g_{dB} = 6$). Audio for all SUTs, including anchors, was level-aligned based on an ITU-R BS.1770 [18] speech-gated measurement [19, 20] to -31 LKFS, a value chosen to avoid clipping. The alignment facilitates direct comparisons between systems and mitigates issues with mutual dialog and background level differences noted in [1].

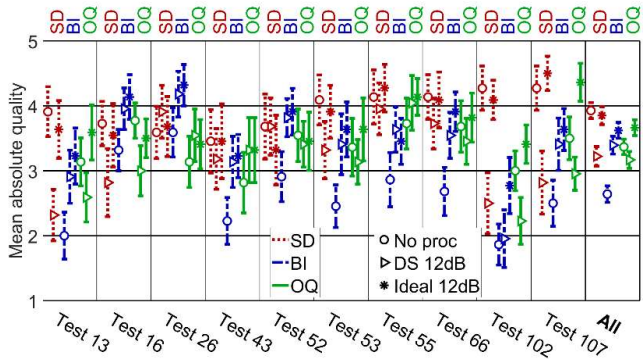

Figure 3: Results for ACR test on the MUSHRA test items

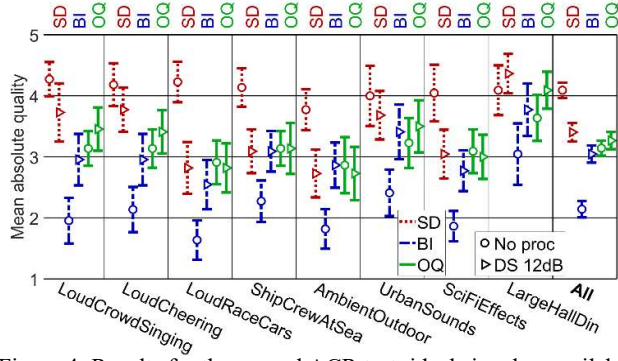

Figure 4: Results for the second ACR test; ideal signal unavailable

MUSHRA results for ten expert listeners are shown in Fig. 2, with legend symbols indicating means across subjects and error bars representing 95% confidence intervals on the means. Out of range of the plot are the hidden references which listeners always scored 100, and the lowpass 3.5 kHz anchors which listeners scored between 9 and 25. We see that, based on mean values, the DeepSpace system outperformed the other SUTs and anchors overall, though it performed slightly less well than U-NetFB on three items; on two of those items it also narrowly underperformed NSNet2. The UNet-FB+C system, which could be viewed as conceptually similar to the external tested systems in that it exploited only source cues, performed similarly to NSNet2 overall and better than DPCRN. Pilot listeners described artifacts for the source-cue-only systems which led to their often-lower scores: spatial instability and backgrounds pumping in rhythm with speech syllables.

### 3.2 Absolute Category Rating Tests

We have also performed absolute category rating (ACR) testing using a methodology closely related to ITU-T Recommendation P.835 [21]. ACR testing allows use of content where an ideal reference signal is unavailable, and here allows for more independent evaluation of the original, unprocessed signals as they are not directly compared against an ideal reference dialog-boosted signal. As in [21], listeners are asked to evaluate absolute quality of a single condition at a time in three aspects of quality: speech distortion (SD; 1 very distorted to 5 not distorted), background intrusiveness (BI; 1 very intrusive to 5 not noticeable), and overall quality (OQ; 1 bad to 5 excellent), each on an integer-only scale; a higher score is better. Items and conditions were presented in a constrained random order, ensuring that for a test with N items, N-1 other items are presented in between presentations of item 1 etc.

Our methodology deviates from [21] mainly in the instructions on how to score BI. In [21], listeners are instructed to pay attention to only the background when evaluating BI. For our application, we include this text in the instructions pertaining to ability to hear and understand, as this is a key goal of DE [3]: *"Use scores 1, 2 or 3 to discriminate between different levels of difficulty to hear what is being said. Use scores 3, 4 or 5 to discriminate between different levels of background loudness in cases where the background does not interfere with your ability to hear what is being said."* Also, the instructions for OQ deviate somewhat from [21]; listeners are asked to take into account any background distortion, in addition to weighing together the scores of SD and BI.

The first ACR test used all ten items from the MUSHRA test and three conditions: ideal 12 dB boost, DeepSpace 12 dB boost, and unprocessed. Results for 22 expert listeners are shown in Fig. 3. None of these listeners participated in the MUSHRA test. We see that overall, the DeepSpace version gains about as many points on BI over unprocessed as it loses on SD versus the ideal and unprocessed signals (0.75 and .70 point respectively), and has nearly identical OQ scores as the unprocessed original, professionally generated content signal, while missing the ideal boost OQ by about 0.5 points. There are substantial variations between items.

Considering the average DeepSpace results for each item on the MUSHRA test, in comparison with the average SD, BI and OQ scores for DeepSpace for those same items on the ACR test, the respective Pearson correlation coefficients are (.89, .85,.81) with respective *p*-values of (.001, .002, .004). Here, we use Pearson rather than Spearman's rank correlation due to four items with similar MUSHRA scores and strongly overlapping error bars. Comparing the MUSHRA scores with the *difference* between DeepSpace and "Ideal" ACR scores, the coefficients and *p*-values are (.83, .63, .77) and (.003, .049, .010). We interpret the high correlation values of the first set of correlations (and that the values are higher than the second set) as suggesting the possibility that an ACR test could provide most of the same information as a MUSHRA test for DeepSpace, without requiring an ideal reference. The analogous correlations for the unprocessed scores were (.04, -.05, .27) and (.12, -.29, .29) with *p*-values (.92, .89, .44) and (.74, .42, .41). We interpret the low magnitude correlations as a consequence of comparing relatively undifferentiated low scores subjects gave to the unprocessed items in the MUSHRA test (due to their large differences from the reference) to the substantially differentiated, sometimes relatively high ACR test scores.

The second ACR test includes 8 professionally mixed items containing scripted TV and movie content and a wider range of mixing types. Ideal *dialog-boosted signals* are not available. The two test conditions are DeepSpace 12 dB boost and unprocessed. The content includes live sports (items 1-2), motorsports (item 3), scripted TV (items 4-6) and movies (items 7-8). Item 2 has off-center panned dialog, item 6 has various dialog sources each with their own panning, item 8 has reverberant dialog, and the others have only center-panned dialog. Item 4 has male and female dialog, item 7 female-only dialog, and the others male-only dialog. The content names shown in the figure describe the backgrounds for each, with "loud" backgrounds indicating low SNR (under approximately 5 dB); other SNRs were moderate (approximately 5-10 dB). Results, for the same 22 expert listeners, are shown in Fig. 4. We observe that listeners overall gave a lower rating for SD by slightly less than they gave a higher rating for BI (0.7 and 0.9 point respectively) while rating OQ similar; also here there is substantial per-item variation. OQ for DeepSpace was higher on 5 items; no item's OQ average score for DeepSpace was more than 0.2 points below the original, even when SD was lower by 0.5 points or more, indicating that BI also has strong influence. We interpret these results overall as suggesting that the 12 dB DeepSpace dialog boosted signals present a viable alternative to the original, professionally generated mixes.

## 4. AUTOMATED METRICS

Automated metrics can be valuable for system evaluation, as they can process a much larger number of items and SUTs than is practical for human listening. Yet there are no widely accepted metrics which evaluate the specific dialog-boosted signals output by DE systems. Given the potential advantages of automated metrics, we explore the feasibility of using two types of metrics to evaluate DE systems in the next subsection. In a second subsection we explore the viability of DeepSpace and U-NetFB as SE systems based on data from automated SE metrics.

### 4.1 Feasibility of PEAQ and Dialog Estimate Metrics for DE

First, we consider use of the automated metric PEAQ, originally described in [22] and implemented as Matlab code in [23]. PEAQ compares SUTs to an ideal reference, as does a MUSHRA test. Note, however, that PEAQ was trained to evaluate audio codecs whose errors may differ substantially from those surfaced in the DE MUSHRA test above. PEAQ outputs *Objective Difference Grade* values from -4 (very annoying) to 0 (imperceptible) inspired by BS.1116 [24]. Excluding the trivial ideal reference scores, we calculated Spearman's Rank Correlation Coefficient (SRCC) between each SUT's average values for subjective MUSHRA results and PEAQ as -.267 with a *p*-value of .48, suggesting no meaningful relationship.

We also considered the applicability of SE metrics which operate directly on dialog estimates. We consider intrusive (ground truth required) metrics PESQ [25], STOI [26], and ViSQOL [27], and the non-intrusive (no ground truth required) metrics of DNSMOS [28], a DNN-based SE metric. Its component metrics are OVRL which predicts (human rating of) overall quality, SIG which predicts speech quality, and BAK which predicts background intrusiveness, all based on P.835 testing [21].

We described in the introduction that DE and SE differ with regard to intentionally included backgrounds in outputs, input and output bandwidth, and the spatial and channel characteristics of inputs, all of which suggest that these metrics may *not* be applicable to DE. A dialog estimate which is spatially inaccurate and bandlimited might score highly on an SE metric and produce a poor DE experience, while a dialog estimate which has artifacts later to be masked by the (intentionally included) backgrounds might score poorly but provide an excellent DE experience.

Using the SE metrics as designed (on dialog estimate signals, using clean dialog references when applicable) and treating each stereo channel as a test signal, we ran the SE metrics on all true SUTs in the MUSHRA test (the anchors lack defined dialog estimates) and calculated the SRCC values for each metric (STOI, PESQ, ViSQOL, OVRL, SIG, BANK) against the MUSHRA scores per SUT. We obtained the values (-.10, -.30, -.30, -.10, -.10, 0.0) with *p*-values (.95, .68, .68, .95, .95, 1.0), which suggests that none of these metrics have a meaningful relationship with the MUSHRA data.

### 4.2 Speech Enhancement Evaluation

Though the dialog estimate metrics in the previous subsections appear to be ineffective predictors of relative quality for *dialog-boosted* signals, this does not invalidate SE metrics used for their intended purpose. Based on their SE metric scores on pilot tests, we now consider the performance of DeepSpace and U-NetFB as SE systems, using SE metrics. We also consider the above-mentioned SE systems NSNet 2, DPCRN, and FullSubnet.

We ran the noted SE metrics on SUT results for a larger stereo data set, similar to the MUSHRA set, containing clips of TV and movie content for which we also have ground truth dialog. These were obtained from broadcast signals using a process described in [1]. The data set comprises 384 stereo clips, each 10s in duration. The clips include three versions of each of 128 items, with DNRs of 0, 5, and 10 dB based on detected LKFS values of dialog and backgrounds. In order to evaluate system robustness to non-center panning, we mixed speech in a diverse way, including 60% center panned, 30% non-center panned (with a uniformly distributed random panning direction between full left and full right), and 10% moving speech with uniformly distributed random start and end positions and for each item, a single, constant rate of panning change.

We tested each stereo signal as two mono signals, downsampled to 16kHz. Average results for each SUT are shown in Table 1. It can be seen that the U-NetFB system performed best among these systems on STOI, PESQ, and ViSQOL; DeepSpace outperformed only NSNet2. For DNSMOS, DeepSpace achieved the highest score of the SUTs on two of three DNSMOS metrics. These results lead us to consider using U-NetFB and DeepSpace as SE systems for 16 kHz or 48 kHz stereo content in the future.

| SUT | STOI | PESQ | ViS. | OVRL | SIG | BAK |
|---|---|---|---|---|---|---|
| FullSubn. | 0.82 | 2.79 | 2.92 | 2.84 | **3.28** | 3.64 |
| DPCRN | 0.83 | 2.78 | 3.00 | 2.85 | 3.26 | 3.69 |
| NSNet2 | 0.77 | 2.45 | 2.57 | 2.54 | 2.94 | 3.56 |
| U-NetFB | **0.85** | **2.91** | **3.01** | 2.82 | 3.20 | 3.75 |
| DeepSpace | 0.80 | 2.61 | 2.69 | **2.86** | 3.18 | **3.90** |

Table 1: SE metric scores on larger data set.

## 5. CONCLUSION AND FUTURE WORK

We described and presented results for an unguided DE system using DeepSpace source separation, which uses both dynamic spatial cues and source cues. On a MUSHRA test, DeepSpace outperformed other systems available for testing, including those making less advanced use of spatial and source cues and those which did not use spatial cues. On ACR tests, DeepSpace DE provided overall quality similar to that of original, professionally generated content while offering trade-offs on background intrusiveness and speech distortion when testing at a relatively high boost level of 12 dB. We found that automated metrics developed to evaluate codecs or SE systems were not effective predictors of DE performance as subjectively tested, but that DeepSpace and U-NetFB may be viable SE systems for stereo content. In future work, we will explore additional subjective testing methodologies for DE signals and describe a version of DeepSpace optimized for 5.1 content.

## 6. ACKNOWLEDGEMENTS

The authors thank Hannes Muesch, Heiko Purnhagen, Xiaoyu Liu, and Przemyslaw Wojciga for their assistance.